\documentclass[12pt]{article}
                              \usepackage {subeqn,cite,psfig}
\def\be{\begin{equation}}
\def\ee{\end{equation}}
                              \def\barr{\begin{array}}
                              \def\earr{\end{array}}
\def\dis{\displaystyle}
\def\eg{{\em e.g.}}

\def\etal{{\em et al.}}
\def\ie{{\em i.e.}}

                              \def\gev{\: {\rm GeV} }
                              \def\tev{\: {\rm TeV} }
                              \def\pb{\: {\rm pb}}
                              
\def\ra{\rightarrow}

                              \def\ptsl{p_T \hspace{-1.1em}/\;}

\def \rp {$R_p \hspace{-1em}/\;\:$}
\def \st {{\tilde t} }
\def \sb {{\tilde b} }
\def \ch {{\tilde \chi^+} }
\def \nt {{\tilde \chi^0} }
\def\simlt{\stackrel{<}{{}_\sim}}
\def\simgt{\stackrel{>}{{}_\sim}}

\newcommand{\tb}{\tan\beta}

\newcommand{\beq}{\vspace{2mm}\begin{eqnarray}}
\newcommand{\eeq}{\end{eqnarray}\vspace{2mm}}

\begin{document}
%--------------------------------------------------------------------%
%----------------------------Titlepage-------------------------------%
\setcounter{page}{0}
\thispagestyle{empty}
\renewcommand{\thefootnote}{\fnsymbol{footnote}}

\begin{flushright}
                                                CERN-TH/97-176\\[1.5ex]
                                                FNAL-PUB-97/258-T\\[2ex]
                                       {\large \tt hep-ph/9707458} \\
\end{flushright}

\vskip 25pt

\begin{center}
\advance\baselineskip by 10pt

{\Large\bf 

                   Supersymmetry and Charged Current Events at HERA 
}
\\[2cm]

\advance\baselineskip by -10pt

{\bf 
             M. Carena$^{a,b}$, Debajyoti Choudhury$^a$, \\[1.5ex]
             Sreerup Raychaudhuri$^a$ {\rm and} C.E.M. Wagner$^a$ 
}

\rm
\vspace{13pt}

{\em $^a$ 
        Theory Division, CERN, CH 1211 Geneva 23, 
        Switzerland
}

{\em $^b$  
            Fermi National Accelerator Laboratory, 
            P.O. Box 500, Batavia, IL 60510, U.S.A.
}\\[2ex]

E-mails: 
            {\tt mcarena@fnal.gov; debchou,sreerup,cwagner@mail.cern.ch}

\vspace{30pt}

{\bf 
                                 Abstract
}

\end{center}

\begin{quotation}

A light stop, with an $R$-parity-violating coupling
$\lambda'_{131}$, has been suggested as an explanation of the excess in
high-$Q^2$ neutral current events observed at the HERA collider.  We
show that in this scheme a corresponding excess in charged current
events --- such as that reported by the H1 Collaboration --- can
appear naturally, without calling for the presence of light sleptons
or additional $R$-parity-violating couplings, if there exists a
chargino lighter than the stop. The predicted event shapes agree well
with the data. The relevant region of parameter space is identified,
taking into account constraints coming from precision electroweak
measurements, atomic parity violation and recent searches for 
first-generation leptoquarks at the Tevatron collider.

\end{quotation}

\vspace{5ex}

\noindent
                            CERN-TH/97-176\\
                              July 1997
\vfill
\newpage

%-----------------------------Cover page over------------------------%
\setcounter{footnote}{0}
\renewcommand{\thefootnote}{\arabic{footnote}}
\setcounter{page}{1}
\pagestyle{plain}
\advance \parskip by 10pt
\pagenumbering{arabic}
%--------------------------------------------------------------------%

The reported~\cite{H1,ZEUS} excess of neutral-current (NC) events at
large $Q^2$ in deep inelastic scattering (DIS) at the HERA collider
has been interpreted as a possible hint of physics beyond the
Standard Model (SM)~\cite{Adler, ChRa, CERN_gang,  DrMo, DESY_gang,
Blum, BKMW, HeRi, BCHZ, AlGiMa}. Among the various solutions
considered, the currently favoured ones involve the $s$-channel
production of a scalar `leptoquark', which couples to the positron and
one of the valence (or sea) quarks in the proton. Scalars of this
nature, namely {\em squarks}, form an essential ingredient of
$R$-parity violating (\rp) supersymmetric theories.  
Supersymmetry (SUSY) not only provides one of the most well-motivated 
extensions of the SM but also leads to an explanation 
of the observed NC
excess~\cite{ChRa, CERN_gang, DrMo, DESY_gang, AlGiMa}. 
Its role as a framework for solutions to the HERA anomaly thus merits
serious consideration.

Within the framework of the minimal supersymmetric extension of the
Standard Model (MSSM) the relevant \rp\ contribution to the
superpotential reads
\begin{equation}
               {\cal W} = \lambda'_{ijk} L_i Q_j D_k^c  \ ,
       \label{superpot}
\end{equation}
where $i,j$ and $k$ are generation indices, $L_i$ and $Q_j$ are the 
$SU(2)_L$ doublet lepton and quark superfields respectively, while
$D_k^c$ is a charge-conjugate right-handed down quark superfield.  In
view of the various constraints on \rp\ couplings  from low-energy
processes~\cite{Low_ener}, 
the most natural interpretation of the HERA NC data is
given by the $s$-channel production~\cite{Old_HERA}
of a left-handed charm ($\tilde
c_L$) or top ($\tilde t_L$) squark, with a mass of ${\cal O}(200)$
GeV and a coupling to the {\em valence} $d$-quark,
\be
\lambda'_{1j1} \sim \frac{0.04}{\sqrt{\beta_{ed}}},  \qquad
\beta_{ed} \equiv {\rm Br}(\widetilde{q_j} \ra e^+ d) \ ,
                                                    \label{lambda}
\ee
where $j = 2,3$. It is perhaps worth mentioning that a sizeable
effect proceeding from the interaction of the positron with the {\em
sea} quarks in the proton would, in general, demand rather large
couplings, which are generally inconsistent with constraints from
low-energy data. The only exception is the possibility
\cite{CERN_gang, DrMo} of stop production from a sea $s$-quark
through $\lambda'_{132} \simeq 0.4/\sqrt{\beta_{ed}}$.
The parameter space in this scenario has subsequently 
been shown to be constrained by LEP measurements~\cite{ElLoSr}
as well as by limits on the electron-neutrino mass~\cite{JoRaVe}.
In the following, we shall concentrate on a valence quark
interpretation of the HERA anomaly.

Searches at the Tevatron collider for generic leptoquarks decaying
into lepton and jet---of which squarks of \rp\ supersymmetry are
one example---put strong restrictions on the allowed branching
ratio $\beta_{ed}$. For a squark mass of ${\cal O}(200)$ GeV, the
bounds from the CDF~\cite{CDF_LQ} and D0~\cite{D0_LQ} analyses can be
combined to indicate
\be
                    \beta_{ed} \simlt 0.5 \ .
                                                \label{bound}
\ee
While this result seems to rule out most explanations of the HERA
anomaly involving leptoquark 
resonances\footnote{Attempts have been made, however, to construct 
                   models that evade this 
                   constraint~\protect\cite{BaKoRu}.},
which have no other decay
channels, supersymmetric models can easily get around the impasse.
Given the rich particle spectrum of SUSY models, $R$-parity-{\em
conserving} interactions can induce additional decay processes (not
counting, of course, those that lead to hard charged leptons plus
jets in the final state).  If this is indeed the case, with the
negative results at the Tevatron being due to an
$R$-parity-conserving decay that is unobservable with the 
current data sample~\cite{GuRo}, then it is natural to ask what is
predicted at HERA, where a major fraction ($1 - \beta_{ed}$) of the
produced squarks will decay through these modes. One should typically 
expect some additional signals beyond those 
expected within the SM.

As a matter of fact, the H1 Collaboration~\cite{H1}---and 
recently, the ZEUS Collaboration~\cite{ZEUS_new} as well---{\em have}
reported a small excess in charged current (CC) events at large $Q^2$
in the same data sample that contains the NC excess.  Although less
significant statistically than the latter, the observed number is
nevertheless too large to be consistent with SM background
calculations at 2.5 standard deviations.  In view of the expectation
that $R$-parity-conserving decays of the produced squark would be
observed at HERA, an identification of the CC excess with these modes
seems to be the most natural conclusion. It is thus interesting to
ask whether SUSY scenarios can admit this possibility without
contradicting other known phenomenological results. Such an
investigation assumes particular importance since it is 
difficult~\cite{AlGiMa} to
accommodate the CC excess within other non-supersymmetric
explanations for the NC events. Some attempts in this
direction have already appeared in the
literature~\cite{ChRa,AlGiMa,Kim,KonNew}. In Ref.~\cite{ChRa}, it was
pointed out that $R$-parity-conserving decays of the squark could
possibly lead to the CC signal. However, the observed event shape ---
low jet multiplicity, large missing momentum, relatively large
hadronic invariant mass of the final state, and, of course, a high
$Q^2$ --- need to be taken into account in any serious pursuit of
such an explanation. A more detailed study of the event shape has, in
fact, been undertaken, mainly in the context of charm-squark
production, in 
Ref.~\cite{AlGiMa}, where an explanation of the CC excess is obtained
by postulating a light slepton-sneutrino pair, each of which decays
into jets through \rp\ couplings. In order to obtain a signal
consistent with the data, especially with the low jet
multiplicity\footnote{Although, even then, the CC events would, in
            general, tend to have two visible jets (see Fig.2 of
            Ref.~\cite{AlGiMa}).}, 
the sneutrinos must be rather light,
$m_{\tilde{\nu}} \simlt 80$ GeV, while the wino mass $M_2$ should not
be much larger than 200 GeV. 
Similar requirements lead to a
        CC signal consistent with the experimental data in the context of
        top-squark production\cite{AlGiMa,Kim}.
Although this is 
a plausible explanation of the HERA anomaly, light
sneutrinos and low values of $M_2$ lead to a sizeable sneutrino
pair-production cross section at LEP2~\cite{sneut_LEP} and
consequently to observable rates for four-jet final states~\cite{CGLW}. 
This last prediction presents a
knotty problem, in view of the difference between results presented
by the ALEPH Collaboration and the other LEP
experiments~\cite{4-jet}, but the general hope is that a definitive 
result may be obtained from the next run of LEP2.  It is also worth
pointing out that the solution suggested in Ref.~\cite{AlGiMa}
may lead, for scharm production, to events with 
a small fraction of wrong-sign $\ell^- +$ jet final
states at HERA, which could be detectable~\cite{H1_old} as more data
accumulate.
A final state with $\mu^+ +$ jets (with or without missing momentum) 
can also be accommodated in their scenario in the presence of 
a light smuon-sneutrino pair. 

In this letter, we describe a scenario in which a CC excess
with the appropriate event shape may be generated even in the absence
of light sleptons. We concentrate on the stop interpretation with an
\rp\ coupling $\lambda'_{131}$ consistent with Eq.
(\ref{lambda}). A light stop in the 200 GeV mass range is natural in
models with radiative breaking of electroweak symmetry where squarks
of the first two generations are heavy~\cite{Cohen}. It is perhaps better
motivated, therefore, than the competing option of a 
light charm-squark. The hierarchy
of masses of the left-handed stop and the sbottom arising from the
$SU(2)$-breaking condition, 
\be
         m^2_{\tilde t_L} -  m^2_{\tilde b_L}
       = m_t^2 - m_b^2  + m_W^2 \cos 2 \beta  
       > 0  \ ,
                                            \label{sq_mass_split}
\ee
suggests a simple way of generating CC events: if an 
$R$-parity-conserving decay mode of the stop leads to 
sbottoms in the final
state, the CC events may be generated by the subsequent decay $\sb_L
\ra \bar{\nu} d$ of the sbottom through the same \rp\ coupling
$\lambda'_{131}$. (A similar scenario for the $\tilde c_L$ is not
possible, since the $\tilde s_L$ would be somewhat heavier than the
$\tilde c_L$ for $\tb > 1$.)

In principle, if the sbottom is light enough ($m_\sb \simlt 120$
GeV), a substantial branching fraction for the weak decay of a stop
into a bottom squark and a $W^+$ (with the last going mainly into
jets) may be allowed. This has been proposed recently by Kon \etal
\cite{KonNew} as the cause of the CC excess at HERA.  Although 
$\st \ra \sb + W^+ \ra \ptsl\ + $ {\rm jets}  is a straightforward and
economical scenario, it has several potential problems. First, it
would mostly lead to CC events with a jet multiplicity of {\em two}
or {\em three}, whereas, of the four events observed by H1, only one is
consistent with more than a single jet~\cite{Sirois}. Moreover, the
hadronic mass distribution is rather broad and one has to appeal to
the low statistics~\cite{KonNew} to claim any agreement with the
data. Apart from such kinematic arguments, one also encounters the
indubitable fact that the contribution of the stop--sbottom doublet to
the $\rho$ parameter tends to be too large ($\Delta\rho_\st \simgt 2
\times 10^{-3}$) if the stop--sbottom mass splitting is so large. 
Slightly smaller values of $\Delta\rho$ may be achieved by
introducing large mixing angles in the stop and sbottom sectors, and
tuning them appropriately. This has, in fact, been invoked by the
authors of Ref.~\cite{KonNew}. However, even then,
$\Delta\rho_\st \simgt 1.5 \times 10^{-3}$ and hence the scenario
is only marginally consistent at the $2 \sigma$ level~\cite{Alta}. 
In addition, constraints from atomic parity-violation experiments tend
to be more severe in the case of large stop mixing~\cite{Strumia}.

All these difficulties can, of course, be overcome trivially by
pushing the sbottom mass to larger values. This would immediately
make the jets coming from the $W^+$ particle softer, and hence less
often visible, while increasing the missing $p_T$ and the invariant
hadronic mass of the CC events. It also reduces the contribution to
the $\rho$ parameter. Unfortunately, when the sbottom mass is pushed
high enough to reduce the contribution to the $\rho$ parameter to
acceptable values, the $R$-parity-conserving branching fraction
becomes very much smaller than the \rp\ one, and the
corresponding CC event rate gets highly suppressed.

It is thus clear that a phenomenologically consistent explanation for
the CC observation will require light supersymmetric particles in
addition to the $(\st,\sb)$ pair.  The light slepton option has been
commented on above and will not be looked into any further.  Since the stop
decay into neutralinos is naturally 
suppressed\footnote{One could consider a non-negligible mixing 
          between stop and the left-handed charm squark, and 
          thus allow the tree-level decay $\st_L \ra c \nt $. It can be
          shown, however, that the mixing angle must be very large 
          ($\sim 1$) to lead to the observed signal; such a large 
          mixing angle is difficult to accommodate within any 
          reasonable theoretical framework.}, 
the obvious course is to have a {\em chargino} lighter than
the stop. For most of the MSSM parameter space, this also means a
neutralino that is considerably lighter and is, therefore, the
lightest supersymmetric particle (LSP).  Assuming that there are no
other light sfermions, just three decay channels are now open to the
chargino, namely into the neutralino ($\ch \ra \nt W^* \ra \nt f \bar
f'$), into the sbottom ($\ch \ra  c \sb$)
through the Cabibbo--Kobayashi--Maskawa (CKM) mixing (provided the
sbottom is lighter than the chargino), and an \rp\ decay 
through a virtual stop. For most of the parameter space of interest, 
however, the \rp\ decay of the chargino has very small width.
 We shall assume, in the
following, that the $\st$--$\sb$ mass splitting is $\simlt 35$ GeV, 
which seems optimum for the $\rho$-parameter constraint, and that 
$m_\st > m_\ch > m_\sb$. Now, depending on the branching ratios, 
two distinct possibilities present themselves.

($i$) {\em The chargino decays predominantly into the neutralino.} If
the latter is lighter than the sbottom ($m_\nt < m_\sb$), it can 
have only a three-body decay ($\nt \ra b \bar{d} \nu_e, \: \bar{b} d
\bar \nu_e$). In this case, the final state would typically have
multijets and relatively low missing momentum. Such configurations
will probably be lost in the DIS (CC) background coming from SM
interactions. On the other hand, if the neutralino is heavier than
the sbottom (which thus becomes the LSP), \ie
\[
                   m_\st > m_\ch > m_\nt > m_\sb  \ , 
\]
then it would decay into $b + \sb_L$ and the \rp\ 
two-body (and only) decay of this $\sb_L$ would result in a
significant amount of missing momentum. Although, at  first sight,
it might seem that we would still have too many jets, most of these
are now very soft as long as $m_\st - m_\sb \simlt 35 \gev$. A simple
parton-level Monte Carlo simulation shows that most of the jets (apart
from the one coming from the $\sb_L$ decay) fail to meet the minimum
energy criteria of the H1 experiment for defining a
jet~\cite{H1,Sirois}---this typically results in 
most events ending up with only a
single `jet' and almost none with more than two. Even when the parton
energy is adequate, the second jet would be soft and broad, and it
is debatable whether it would be detectable at all. It is quite easy
to confirm that the kinematic profiles for these configurations match
those for the observed CC events. Thus, a cascade decay
of stop into chargino into neutralino into sbottom seems to be an
attractive scenario. However, like the ones considered 
above, this too suffers
from an inherent drawback. Since {\em all} the charginos produced in
the stop decay (and we must remember that this is the dominant decay
of the stop for $\beta_{ed} \simlt 0.5$) are potential CC event
candidates (modulo detection efficiency), we now tend to get too many
of the latter.  Unless one wishes to appeal to experimental effects
and the low statistics, this number can be reduced only by providing
the sbottom with another decay channel. What can that be? If we
violate gaugino mass unification and choose the soft
supersymmetry-breaking parameters such that the mass of the lightest
neutralino ($\nt_1$) is well below that of the sbottom while the
next-to-lightest neutralino ($\nt_2$) lies just below the chargino
and is the one involved in the above chain, then, indeed, a new decay
mode $\sb_L \ra b \nt_1$ can be opened.  A scan of the
parameter space shows that this rather complicated scenario can be
achieved for a limited part of the parameter space and that
too, only if there are large deviations from gaugino mass
universality. Though this scheme works --- as we have checked ---
it has limited aesthetic appeal and will not concern us any further.

($ii$) {\em The chargino decays predominantly into the sbottom and a
charm quark.}
This mode is naturally suppressed by the smallness of
the CKM matrix element $V_{cb} = 0.036-0.046$.
Although some enhancement can be obtained in the event of 
a nonzero $\sb$--$\tilde s$ mixing, we prefer to be conservative 
and do not consider this option. Thus for $\ch \ra
c\sb_L$  to be the dominant decay mode, the chargino--neutralino
coupling which drives the competing decay ($\ch \ra \nt ff'$) has to
be quite small. Fortunately, this occurs naturally when
$|\mu| $ is large compared to $M_2$. The lightest neutralino is then mostly
bino-like and couples very weakly to the $W$, while the lighter
chargino and the next-to-lightest neutralino are mostly wino-like and
nearly degenerate in mass. Thus, the chargino decay into the second
neutralino will be strongly suppressed, if not disallowed. The mass
hierarchy obtained in this scheme is of the form
\[
        m_\st > m_\ch \simeq m_{\nt_2} > m_\sb > m_{\nt_1}  \ . 
\]
We find that a branching ratio $\beta_{ed} \simlt 0.5$ can naturally
be obtained for values of $M_2 \sim m_\ch \sim 190$ GeV) 
and corresponding values of $\lambda'_{131}$ as
determined by Eq. (\ref{lambda}). In order that the jet resulting
from the charm quark should not be observable, we also require the
sbottom to be fairly heavy, typically in the 170--180 GeV range,
which is already indicated by the $\rho$ parameter constraint.  The \rp\
decay of the sbottom into $\bar \nu d$ now provides a hard jet and missing
momentum, as in the cases discussed above, which can be the
origin of the observed CC excess.

For this last scenario---which will concern us in the rest of this
letter---we observe that the sbottom can also decay in the 
$R$-parity-conserving mode $\sb_L \ra b \nt_1$, with the neutralino then
decaying into 
\be
\nt_1 
\ra 
\; \bar b d \bar\nu_e \ , b \bar d \nu_e \ .
      \label{sb_to_neut}
\ee
through an intermediate sbottom.
If the neutralino is light enough, $m_{\tilde{\chi}_1} \sim 100$--140
GeV, the missing $p_T$ of this decay mode will tend to be low, and
will mostly fail the CC selection criteria imposed by the H1
Collaboration (see below).  Potential CC events resulting from
this decay channel will then be lost in the SM DIS background. 
We can thus
have a reduced number of CC events without making any assumption
beyond that of a large $|\mu|$ and $M_2 \sim 190 \gev$.
The simplicity and economy of this scheme is apparent. In effect, 
we only assume:
\vspace*{-3ex}
\begin{itemize}
\item[A.] a small stop--sbottom mass-splitting, which is easily
achieved with a modicum of left--right mixing in the stop sector; 
\item[B.] a chargino, which is mainly wino-like, lying between the stop
and sbottom in mass; and 
\item[C.] a light neutralino, which is
dominantly bino-like, and whose mass is considerably below that of the
sbottom, typically 100--140 GeV. 
\end{itemize}
With these assumptions, a study of
the predicted event shapes and efficiencies is now in order.

In the presence of left--right mixing ($\theta_\st = 0$ implies 
that the lighter eigenstate is a pure
$\tilde t_L$), the stop production cross section at HERA is given by
\be
     \sigma (m_\st) = \dis \frac{\pi}{16 E_e^0 E_p^0} 
             \left( \lambda'_{131} \cos \theta_\st \right)^2
             \: K \: f_d\left(\frac{m^2_\st}{4 E_e^0 E_p^0} \right)  \ ,
    \label{prodn}
\ee
where $E_e^0, E_p^0$ are, respectively, the initial electron and
proton energies in the laboratory frame and $K \simeq 1.3$
parametrizes the QCD correction~\cite{QCD}. For $f_d(x)$, the
$d$-quark distribution within the proton, we use the
MRS(H)~\cite{MRSH} structure functions, calculated using the package
{\sc pdflib}~\cite{PDFLIB}.  In this analysis, we neglect 
effects due to initial state radiation. 
The NC and CC signals are estimated
using a parton-level Monte Carlo event generator. 
For jets within the angular coverage of their detector, the H1 
Collaboration requires a minimum transverse momentum of
\be
     p_T^{\rm jet} > 15 \gev \ .       
           \label{cut:jet_pT}
\ee
In our parton-level analysis,
each final-state quark is assumed to give rise to jets. All such
`jets' that lie within a cone
\be
      \Delta R_{jj} \equiv \sqrt{ \Delta \eta^2 + \Delta \phi^2} \leq 1.0
         \label{cut:jet_merge}
\ee
of each other are `merged' to form a single jet by the simple
expedient of adding their four-momenta together vectorially and
identifying the sum with the momentum of the combination.  Here
$\Delta \eta$ is the difference of pseudorapidities and $\Delta \phi$
denotes azimuthal separation.

For NC events, the H1 analysis requires that the final-state positron
must lie within 
\be
     10^\circ < \theta_e < 145^\circ  \ ,
           \label{cut:pos_angcov}
\ee       
(the proton direction is positive)
and it must have a minimum transverse energy:
\be
     p_{Te} \geq 25 \gev             \ .
           \label{cut:pos_pT}
\ee
In addition, the missing transverse momentum must satisfy $\ptsl \: /
\sqrt{p_{Te} } \leq 3 \gev^{1/2}$. The DIS Lorentz invariants can be
estimated using either of the pairs ($ \theta_e, \theta_{\rm jet}$)
and ($\theta_e, E_e $); the latter give :
\[  \dis
   y_e = 1 - \frac{E_e}{E_e^0} \sin^2 \theta_e, \quad
   Q_e^2 = \frac{p^2_{Te} }{1 - y_e},  \quad
   M_e   = \sqrt{ \frac{Q_e^2}{y_e} }, 
\]
where $E^0_e = 27.5 \gev$ is the energy of the incoming positron
beam. The NC events are required to satisfy
\be
   0.1 < y_e < 0.9 \ .
           \label{cut:ye}
\ee
With the above set of cuts (and other standard ones required to
remove cosmic-ray backgrounds) the H1 Collaboration~\cite{H1} reports
12 events with $Q_e^2 > 15000 \gev^2$ for an integrated luminosity of
14.2 pb$^{-1}$ against the expected DIS background of $4.7 \pm 0.76$
events.

For the CC events, on the other hand, the cuts
(\ref{cut:pos_angcov})--(\ref{cut:ye}) are no longer operative.
Instead, the H1 analysis demands that the missing transverse momentum
be large:
\be
    \ptsl \: = \sum p_T({\rm hadrons})  > 50 \gev \ .
             \label{cut:missmom}
\ee
Analogous to the NC case, one can define DIS variables~\cite{Jac_Blo}
using only measurable hadronic variables:
\[ \dis
   y_h = \frac{ \sum (E - P_{||}) }{2 E_e^0},  \qquad
   Q_h^2 = \frac{\ptsl^2}{1 - y_h}  \ ,
\]
where the sum runs over all observed hadronic clusters (in our
simulation, `jets' that pass the cuts).  With a selection criterion
of
\be
    y_h < 0.9 \ , 
           \label{cut:yh}
\ee
the H1 Collaboration observes 4 events above $Q_h^2 > 15000 \gev^2$,
whereas the expected SM background is $1.77 \pm 0.87$ events.

It is easy to see that the efficiency for the NC events should be
solely determined by the stop mass $m_\st$, and for $m_\st \sim 200 \gev$,
can be approximated by
\be
    \epsilon_{NC} \approx 0.56 \ ,
              \label{NC_eff}
\ee
where we assume detector efficiency to be unity.  The CC
efficiencies, on the other hand, {\em can} depend on all the relevant
masses in the decay chain. Again, our computations show that the only
significant dependence is on the bottom-squark mass, and
\be
    \epsilon_{CC} \approx 0.37 \:(0.45) \pm 0.008, \qquad 
       {\rm for} \ m_\sb = 170 \: (180) \gev \ .
              \label{CC_eff}
\ee

\begin{figure}[h]
\begin{center}
\vskip 3.95in
      \relax\noindent\hskip -5.0in\relax{\includegraphics{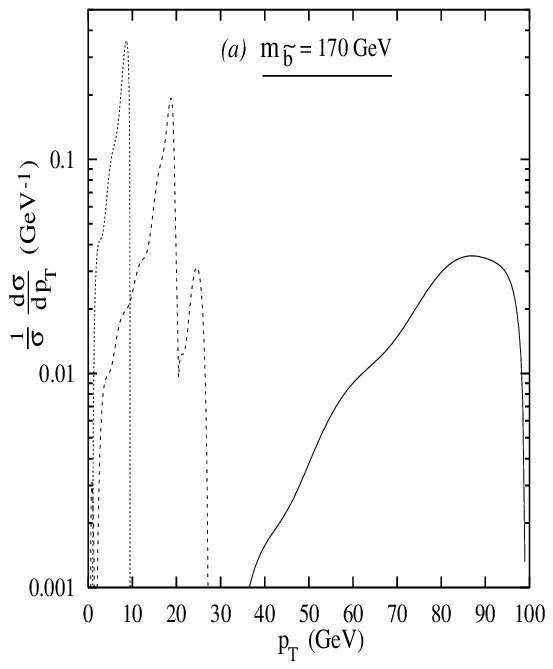}}
      \hskip 2.0in\relax{\includegraphics{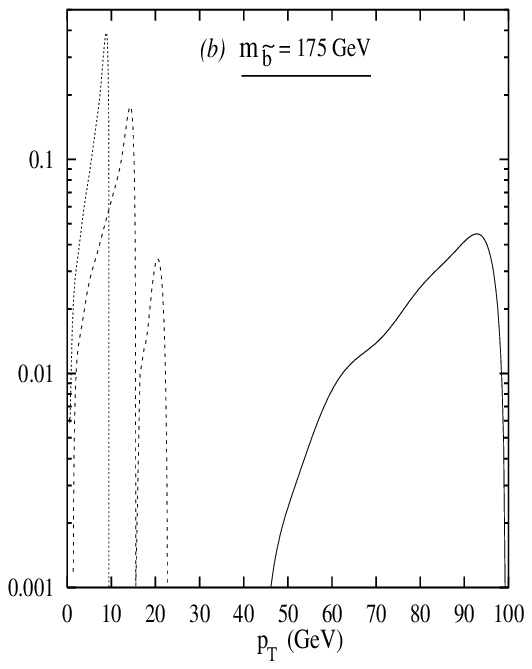}} 
      \hskip 2.0in\relax{\includegraphics{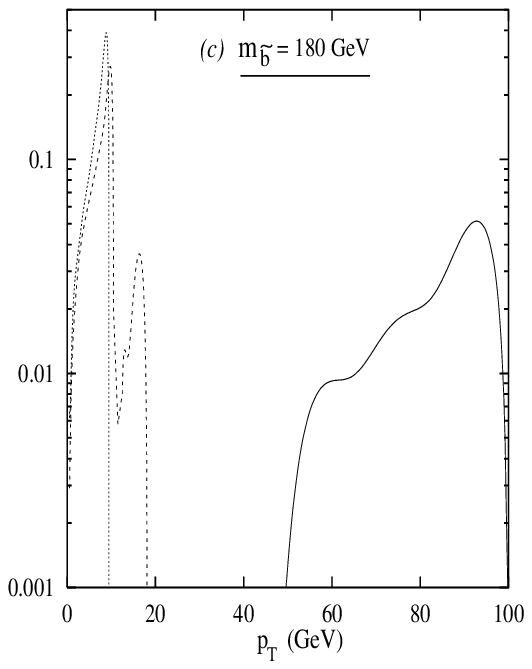}} 
\vskip 2.5in
      \relax\noindent\hskip -6in\relax{\includegraphics{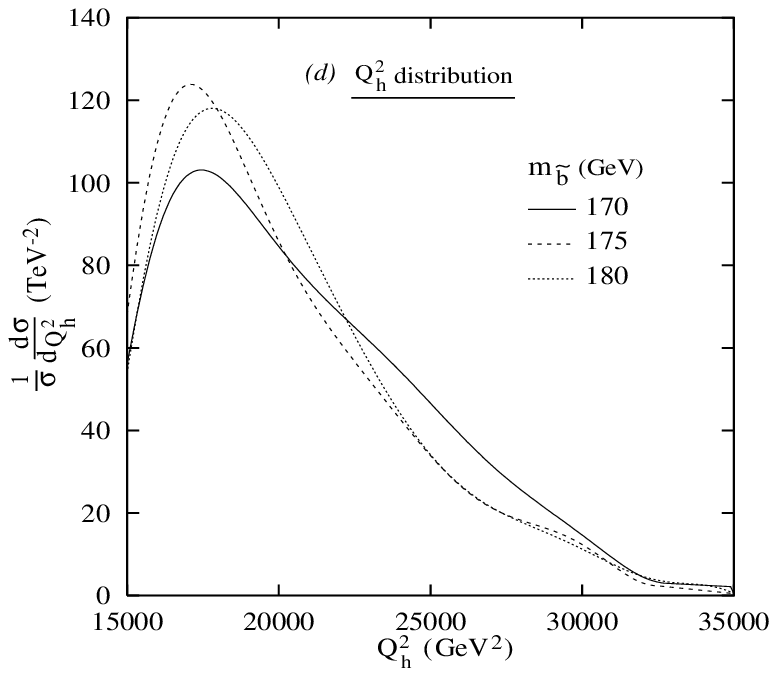}}
      \hskip 3.3in\relax{\includegraphics{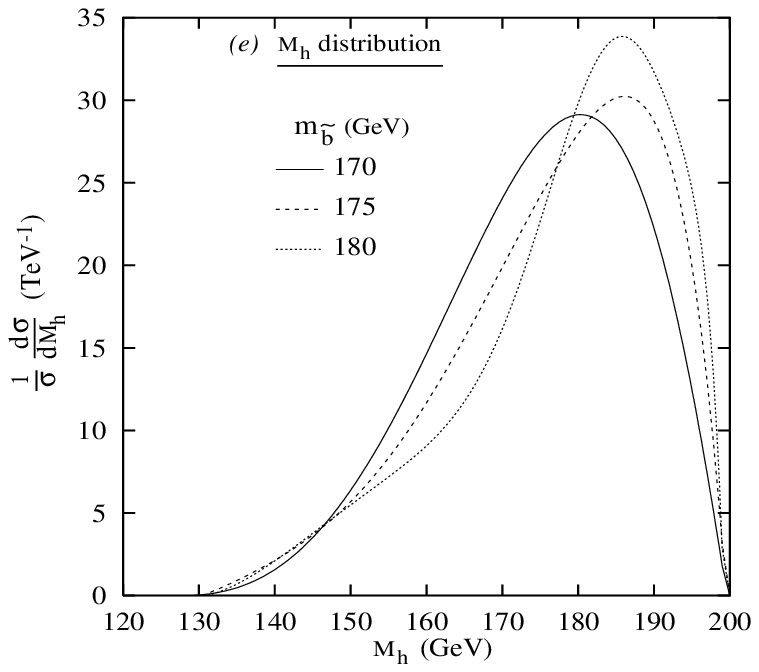}} 
\end{center}
\vspace*{-120pt}
\caption{\em Event topology for 
             $m_\st = 200 \gev$, $m_\ch = 190 \gev$, 
             and for different values of $m_\sb$. 
          {\em (a--c)} The $p_T$ distributions for the three `jets' 
             in each case.  
          {\em (d)} The $Q_h^2$ and 
          {\em (e)} the $M_h$ distributions for all three cases.  
        }
\label{fig:cc_topology}
\end{figure}

As the NC event topology has been studied 
elsewhere~\cite{ChRa, CERN_gang, DrMo, DESY_gang, AlGiMa},
 and shown to be
consistent with the data, we shall not repeat the exercise here. In
Fig.~\ref{fig:cc_topology} we present the kinematic distributions for
the CC events. To be concrete, we have chosen to depict it for
$m_\st = 200 \gev, m_\ch = 190 \gev$ and for three typical sbottom
masses. It turns out that the phase-space distribution 
is not very sensitive to the exact value of $m_\ch$
and is primarily governed by the sbottom mass. 

%%%%%%%%%%%%%%%%%%%%%%%%%%%%%%%%%%%%%%%%%%%%%
%%%%%%%%%%%%%%%%%%%%%%%%%%%%%%%%%%%%%%%%%%%%%
Each of Fig.~\ref{fig:cc_topology}($a$--$c$) depict the $p_T$
distribution of the three potential `jets' for a given $m_\sb$.  As
expected, the hardest jet emanates mainly from the $\sb$-decay, but
with (small) contributions from the soft jets that merged with it.
Consequently, its $p_T$-distribution just reflects the Jacobian
peak with a slight smearing due to the small transverse boosts. 
For the next-to-hardest jet
(dashed lines), the situation is quite different. In general, it has
relatively low $p_T$ and is mostly rejected by the cut
(\ref{cut:jet_pT}). However, depending on the $\st$--$\ch$  and
the $\ch$--$\sb$ mass splittings, some configurations may have 
sufficient $p_T$ to be observable. The remaining jet, if it has 
not merged with the hardest two, is always too soft to be 
seen. This is amply demonstrated by the dotted lines. 
Thus, even though we started with as many as
three quarks in the final state, we never have more than two jets at
the observable level. Moreover, the fraction of 2-jet events is
always much smaller than that of single-jet events for 
$m_\sb \sim 180 \gev$; the
fraction naturally goes up as the sbottom mass decreases. For the
case where two jets can indeed be seen, their angular separation
tends to be evenly distributed within the interval $1 \leq \Delta
R_{jj} \leq 3$.

Figure~\ref{fig:cc_topology}($d$) shows that the number of CC events is
a rather sensitive function of $Q_h^2$; most of the events are
concentrated within $15000 \gev^2 < Q_h^2 < 25000 \gev^2$.
This is well in
agreement with the present HERA data~\cite{H1,ZEUS_new}. 
Events with $Q_h^2 > 30000 \gev^2$~\cite{ZEUS_new} would, 
however, be harder to explain within this scenario.
The mass distribution in
Fig.~\ref{fig:cc_topology}($e$) peaks at about 180--190 GeV, but has
a long tail, which may account for the fact that one of the observed
events has a low $M_h$ ($\sim 157 \gev$). It is interesting that such
a peak in the mass distribution can be obtained in the scenario of
Ref.~\cite{AlGiMa} only for rather small sneutrino masses, while the
scenario of Ref.~\cite{KonNew} has a much flatter and broader mass
distribution.

In Fig.~\ref{fig:cc_topology}
we present results for only a few points in the
available parameter space, but it is easy to see that the same
qualitative features would hold for other points as long as these
satisfy the conditions (A-C) above. What of the events resulting 
from the decay chain of Eq.(\ref{sb_to_neut})? A simulation analogous
to the one described above shows that most of these events indeed 
fail to survive the cuts of 
Eqs.(\ref{cut:jet_pT}, \ref{cut:missmom}--\ref{cut:yh}), 
with the missing momentum cut proving to be the most severe. 
The corresponding efficiency is a sensitive function of $m_\nt$ 
and ranges from about 4\% to 13\% as the latter varies from 
90 GeV to 140 GeV. This is rather too low to yield an observable 
number of multijet events with large missing momentum in the 
present data sample.

\begin{figure}[h]
\begin{center}
\vskip 3.95in
      \relax\noindent\hskip -6.5in\relax{\includegraphics{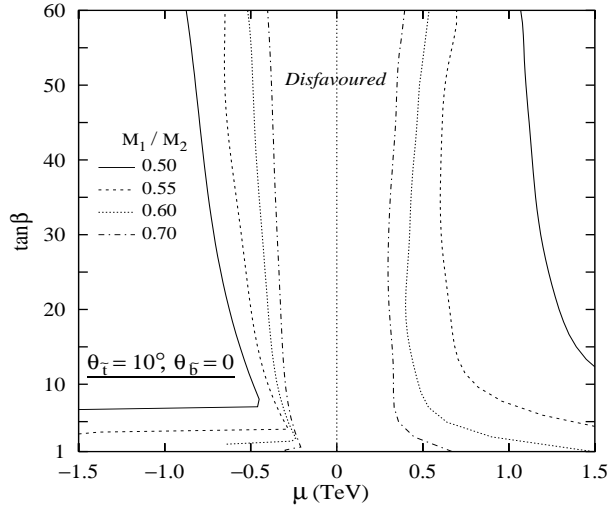}}
\end{center}
\vspace*{-120pt}
\caption{\em The favoured range (see text) in the $\mu$--$\tan \beta$
             plane for 
             $(\lambda'_{131} = 0.07, m_\st = 200 \gev, m_\sb = 175 \gev)$  
             for four values of $M_1/M_2$. We assume a small 
             stop mixing.
        }
        \label{fig:par}
\end{figure}

Having established that the HERA results for CC events can be
explained within the present scenario, it becomes interesting to identify
the part of the MSSM parameter space that supports our solution.
Certain requirements are immediately obvious. As explained above,
since we want the chargino to be wino-like and slightly lighter than
the stop, we are immediately constrained to $M_2 \sim 190 \gev$.
For $\mu$, $\tb$ and the ratio $M_1/M_2$, the situation is more
complicated. We note that the ratio of the number of NC events to CC
events in this scenario is given by
\be \dis
    \frac{{\rm N_{CC}}}{{\rm N_{NC}}} 
          =    \left( \frac{1-\beta_{ed}}{\beta_{ed}}      \right)
            \; {\rm Br}( \ch \ra c \sb)
            \; {\rm Br}( \sb \ra \bar{\nu} d)  
            \; \left( \frac{\epsilon_{CC}}{\epsilon_{NC}}  \right)
                    \ ,
        \label{CC_by_NC}
\ee
with efficiencies as in Eqs.~(\ref{NC_eff}) and (\ref{CC_eff}). 
Consider a typical value $\beta_{ed} \simeq 0.4$ consistent with the
Tevatron constraints. Since
${\rm Br}(\sb \ra \bar{\nu} d) \approx 0.4$~(0.6) for 
$m_\nt \sim 100$~(140) GeV,
 we have 
$N_{CC}/N_{NC} \simeq$ (0.35--0.5) $\times {\rm Br}(\ch \ra c \sb) $. 
A good agreement with the H1 rates thus requires
${\rm Br}(\ch \ra c \sb) \simgt 0.5$. This constrains the parameter
space in the $\mu$--$\tb$ plane, and we present the favoured region
in Fig.~\ref{fig:par}.  While this region depends on the value of the
\rp\ coupling $\lambda'_{131}$ as well as on the squark masses
(these not only determine the production cross section and the
efficiency, but also bound $M_2$), we present the parameter
space for only one representative set of values.  To be deemed
acceptable, a point in the parameter space was required to lead to $7
\pm 1$ NC events and $2.5 \pm 1$ CC events for an integrated
luminosity of $14.2 \pb^{-1}$ and simultaneously 
satisfy  $0.3 \leq \beta_{ed} \leq 0.5$.
Note that these bounds on $\beta_{ed}$
are relaxed for a slightly heavier stop (say 210 GeV).
This would considerably expand the favoured region.
A similar effect can be obtained by considering a mixing 
the $\sb$--$\tilde s$ sector. 
            Our estimate should thus be regarded as a conservative one.

%%%%%%%%%%%%%%%%%%%%%%%%%%%%%%%%%%%%%%%%%%%%%
%%%%%%%%%%%%%%%%%%%%%%%%%%%%%%%%%%%%%%%%%%%%%

As shown by the solid lines in Fig.~\ref{fig:par}, for our scenario
to work within the scheme of universal gaugino masses ($M_1 / M_2 =
\frac{5}{3} \tan^2 \theta_W \simeq 0.5$), a relatively large $|\mu|$ is
indicated. The curious shape of the curves owes its origin to the
existence of minima in the wino content of the LSP as $\tb$ is
varied. The asymmetry between positive and negative $\mu$ can be
traced to a similar source.  While $\mu > 1 \tev$ (or $\mu < -600\ 
\gev$) is perfectly acceptable, it is interesting to ask if smaller
values could be accommodated. An elegant way to achieve this is 
to reduce the mass difference $m_\ch - m_\nt$ so that the two-body
decay $\ch \ra \nt W^+$ is suppressed kinematically and the relevant
process is the {\em three-body} decay of the chargino into the LSP
and two light fermions. This happens naturally in the event of small
corrections to the universality relation in the gaugino sector. In
Fig.~\ref{fig:par}, we also show the favoured region for
three other values of $M_1/M_2$. {\em A priori}, the dependence on 
this ratio could be quite complicated. While a larger value of $M_1/M_2$
would kinematically suppress the $\ch \ra \nt f \bar{f}'$ channels, 
it would also increase the higgsino content of both
the chargino and the LSP and consequently the $\ch \nt W$ 
coupling. At the same time, the $\ch c \sb$ coupling decreases too. 
Still, for the range of interest, the dependence is monotonic. 
It must be noted though that $M_1 / M_2 > 0.7$ implies 
$m_\nt \simgt 140 \gev$, which runs the risk of resulting in 
a few (possibly observable) multijet events 
with large missing momentum. 

In Fig.~\ref{fig:par}, we have chosen particular 
values for the squark mixing. A non-zero value for $\theta_\sb$ will
serve to modify both ${\rm Br}(\ch \ra c \sb)$ and 
${\rm Br}(\sb \ra \bar \nu d)$.
Since the $\sb \bar b \widetilde B$ coupling vanishes exactly 
for $\tan \theta_\sb = 0.5$, for a mixing angle in the vicinity of 
this value, the curves in Fig.~\ref{fig:par} would move {\em inwards} 
by approximately 150 GeV. 
Stop mixing, on the other hand, manifests itself 
mainly in determining the stop-chargino coupling and hence 
$\beta_{ed}$. For $|\theta_\st| \simlt 15^\circ$, the dependence 
on this parameter 
is negligible for negative $\mu$. For positive $\mu$, the 
deviation from Fig.~\ref{fig:par} 
is maximum for $M_1 / M_2 \sim 0.5$ and, for moderate 
$\theta_\st$, could, at best, lead to a shift by $\sim$~100--150 GeV, 
with the weakest constraints being obtained for $\theta_\st \approx 0$.

A further interesting feature of this scenario is the possibility 
of the chargino decaying into a neutralino and a $\ell^+ \nu$ pair 
instead of jets. This would typically lead to a hard lepton, jets 
and missing momentum in the final state. For $\ell = \mu$, this 
could perhaps explain~\cite{KKK} the muonic event~\cite{Muon}
observed by H1. A similar event with  the $\tau^+$ leads 
to a narrow jet and is swamped by the CC DIS signal. 
An $e^+$ with low missing momentum would, similarly, 
be indistinguishable from 
the NC DIS, but events with larger missing momentum might be 
detectable as statistics multiplies.

We now turn to the various low-energy measurements that, in
principle, could further constrain the allowed parameter space.  As
we have seen in the above discussions, 
if the stop is responsible for both the NC and the CC excesses at HERA,
it cannot be a pure left-handed state.  The corrections to the $\rho$
parameter would then be too large to be consistent with precision
electroweak measurement data.  While a large mixing can evade this
problem, such a solution runs counter to the bounds from atomic parity
violation~\cite{Strumia}.  
The effective weak charge measured experimentally now
receives a contribution from both stops; for a nucleus $^{Z}X_A$,
it is given by
\be
\Delta Q_W = -\frac{\left( \lambda'_{131} \cos \theta_\st \right)^2}
                   {2\sqrt{2}G_F} (2 A - Z)
\left[ \frac{1}{m^2_\st} + 
       \frac{\tan^2\theta_\st}{m^2_{\tilde{T}} } \right].
    \label{apv_effect}
\ee
where $\tilde T$ denotes the heavier stop. The experimental
bound~\cite{APV} on $\Delta Q_W$ of the Cesium atom, 
along with the size of the NC
excess at HERA, thus translates into a bound in the ($\theta_\st,
m_{\tilde T}$) plane. For example, $ m_\st \simeq 200$ GeV implies, 
at the 95\% C.L. level, 
\[
   \beta_{ed} \simgt 0.3 \left[1 + \tan^2\theta_\st 
                \;  \frac{ m^2_\st }{ m^2_{\tilde{T}} } \right]
    \ .
\]
It should be remembered though that the error in $\Delta Q_W$ 
is dominated by theoretical uncertainties, and so this 
bound should be considered with great caution. 
It is clear that, given the upper bound on $\beta_{ed}$ imposed by
the Tevatron data, small values of the mixing and/or a large
hierarchy between the stop masses are preferred.

\begin{figure}
 \centerline{
 \psfig{figure=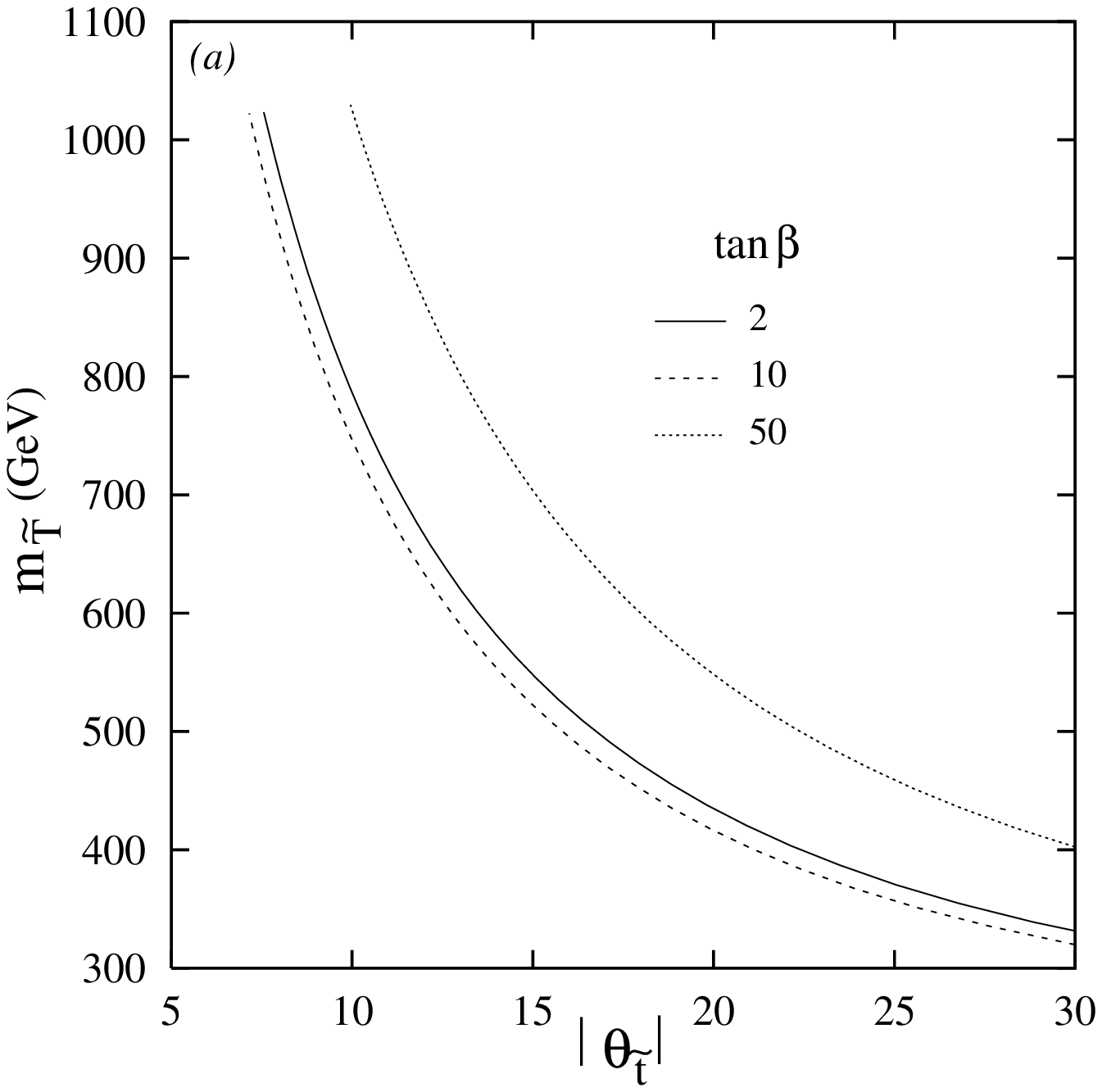,width=8.5cm,height=7cm,angle=0}
 \psfig{figure=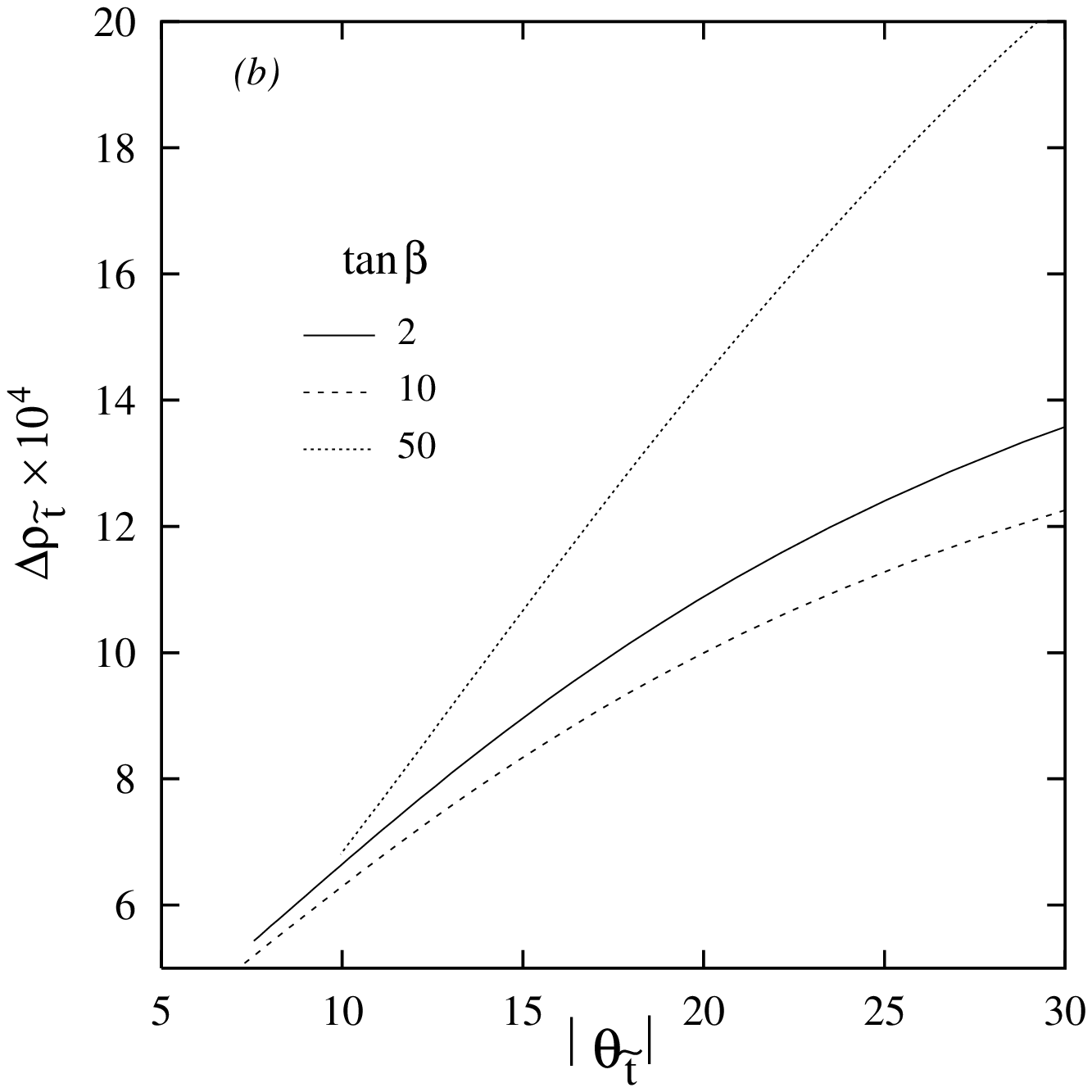,width=8.5cm,height=7cm,angle=0}}
 \caption[0]{\em {\em (a)} The heavier stop mass as a function 
                     of the mixing angle for $m_\st = 200$ GeV,
                     $m_\sb = 175$ GeV, $|\mu| = 1$ TeV, $A_b = 0$, 
                     $m_D = 2$ TeV and $m_U \leq 1$ TeV.
                 {\em (b)} The corresponding contribution of the 
                      stop--sbottom doublet to the  parameter 
                      $\Delta\rho$.
             }
      \label{fig:stop_rho}
 \end{figure}

A simultaneous resolution of the $\rho$-parameter and the atomic
parity-violation constraints thus call for a small $\st$--$\sb$ mass
splitting as well as a small stop mixing angle. While this may seem
difficult at first, in reality, it can arise in a natural way. Consider
the case where $m_U$, the supersymmetry-breaking mass parameter in
the right-handed stop sector,  is large in comparison with that in the 
left-handed sector ($m_Q$).
The lightest stop mass is then approximately given by
\be
m_\st^2 \simeq m_Q^2 + D_L^t + m_t^2 \left(1 -
       \frac{\widetilde{A_t}^2}{m_U^2}\right),
           \label{stop-mass}
\ee
where $\widetilde{A}_t = A_t - \mu/\tan\beta$ and $D^t_L = 0.35 M_Z^2
\cos 2\beta$. The lighter stop will be mainly left handed (\ie\
$\theta_\st$ is small), and the heavier stop mass ($m_{\tilde T} \sim
m_U$) may be pushed to large values. Clearly, the smaller
$\theta_\st$ is, the higher $m_{\tilde T}$ is pushed---see
Fig.~\ref{fig:stop_rho}($a$)---and consequently, the smaller the
additional contribution to $\Delta Q_W$ is.  Ignoring the small mixing
in the sbottom sector, the left-handed sbottom mass will be given by
\be
      m_\sb^2 \simeq m_Q^2 + D_L^b + m_b^2 \ ,
            \label{sb-mass}
\ee
where $D_L^b = -0.42 M_Z^2 \cos 2\beta$. It is clear that, under
these conditions, for values of $|\widetilde{A}_t|   \sim m_U$, the
lightest stop and sbottom masses will be close and hence the
contribution to the $\rho$ parameter will be small. On the other hand, 
for a fixed positive value of $m_\st - m_\sb$, the contribution to the 
$\rho$ parameter is smaller for larger values of $m_U$ and for
$|\widetilde{A}_t|   \sim m_U$.  This conclusion remains valid as 
long as the mixing in the sbottom sector is small. Large $\theta_\sb$
tends to push the value of $m_Q$ up, leading to a substantial mixing
in the stop sector for the same values of the stop masses.
Sbottom mixing
appears naturally in the large $\tan\beta$ region and its effects are
clearly seen in Fig.~\ref{fig:stop_rho}($a$).

%%%%%%%%%%%%%%%%%%%%%%%%%%%%%%%%%%%%%%%%%%%%%%%%%%%%%%%%%%%%
%%%%%%%%%%%%%%%%%%%%%%%%%%%%%%%%%%%%%%%%%%%%%%%%%%%%%%%%%%%%
In Fig.~\ref{fig:stop_rho}($b$), we
plot the stop contribution~\cite{DrHa} to $\Delta\rho$ as a function of
$\theta_\st$ for $m_\st = 200$ GeV and $m_\sb = 175$ GeV.
Fixing $\mu$ and $m_D$, the right-handed sbottom mass parameter, 
allows us to see the effects of a small mixing in the sbottom sector
for large values of $\tan\beta$. Large values of $\mu$ and $\tan\beta$
also have a striking effect on the branching ratio 
${\rm Br}(b \ra s \gamma$) as we shall discuss below. As argued
above, the larger the left-handed component of the lighter stop, the
smaller $\Delta\rho$. 
The experimental numbers suggest~\cite{Alta} 
$\Delta \rho_\st < 1.5 \times 10^{-3}$ 
for $m_t = 175 \gev$ and $m_h \simeq 100 \gev$. From the figure, 
it is apparent that 
$\theta_\st \simlt 15^\circ$ is preferred by the data. Observe that 
even lower values of  $\Delta\rho$ may be obtained by relaxing the 
upper bound on $m_U$.

\begin{figure}
 \centerline{
 \psfig{figure=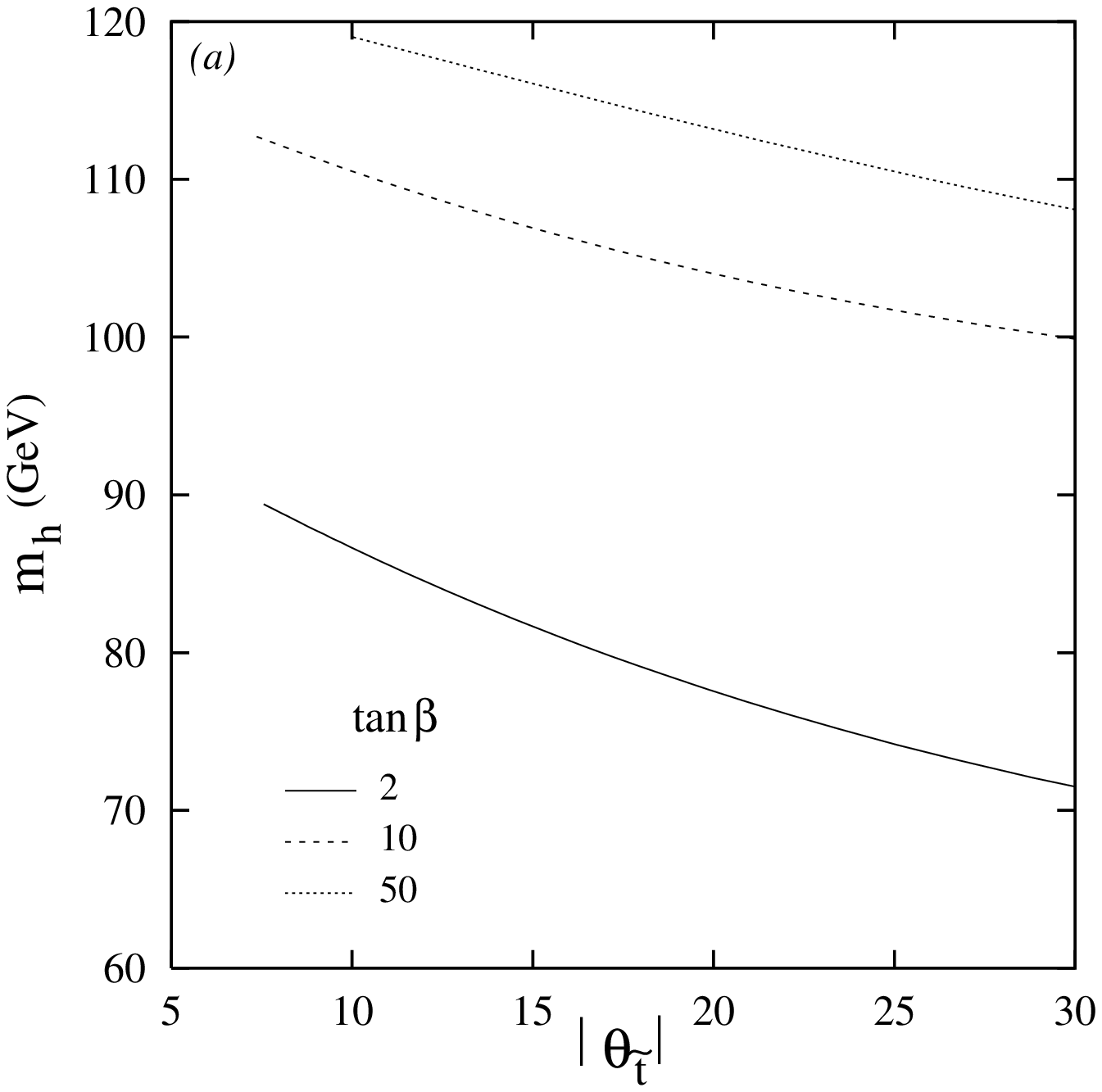,width=8.5cm,height=7cm,angle=0}
%  \hspace*{2em}
 \psfig{figure=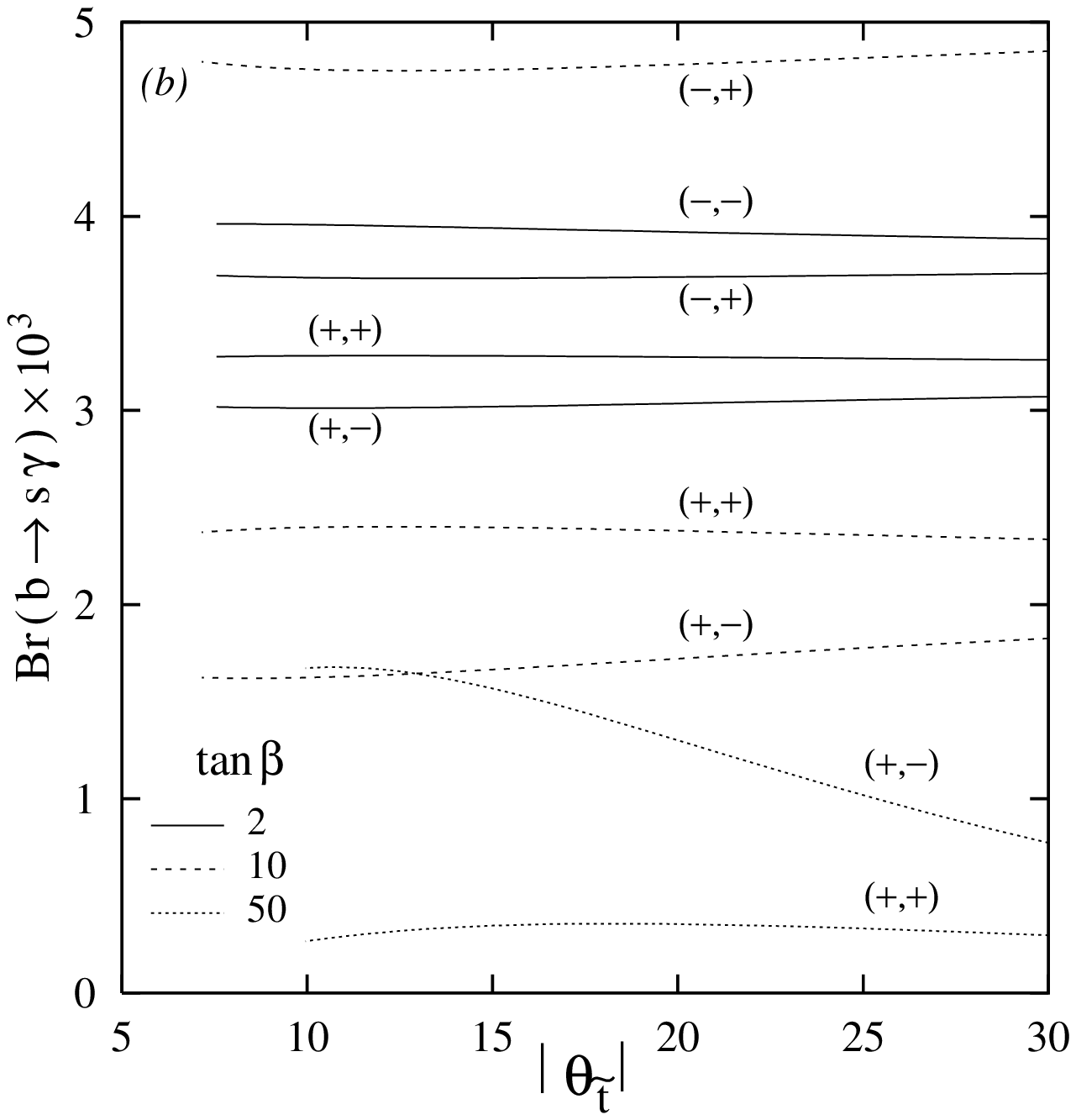,width=8.5cm,height=7cm,angle=0}
  }
\caption[0]{\em {\em (a)} The mass of the lightest CP-even Higgs 
                   as a function of the stop mixing angle, 
                   for the same parameters as in 
                   Fig.\protect\ref{fig:stop_rho} and $m_A = 1 \tev$.
                {\em (b)} {\rm Br}($b\ra s\gamma$) as a function of the
                   stop mixing angle for the same parameters. 
                   The branches are defined by 
                   $({\rm sgn}(\mu), {\rm sgn}(\widetilde A_t))$.
           }
         \label{fig:mh_bsg}
 \end{figure}

Since both stop masses are determined in this scenario, it is
necessary to discuss their impact on the mass of the 
lightest CP-even Higgs~\cite{Higgs_mass}.  
In Fig.\ref{fig:mh_bsg}($a$),
we present the variation of the Higgs
mass as a function of the mixing angle, for the same values of the
stop and sbottom masses. Although there are four branches
associated with the two different signs of $\mu$ and
the mixing parameter $\widetilde A_t$, 
for $m_A \simgt 250 \gev$, there is little difference between the
branches. Hence, we plot the Higgs mass only for positive
values of $\mu$ and $\widetilde A_t$. 
We see that small mixing angles are
preferred for low values of $\tan\beta$, while
 no information on the mixing angle may be 
obtained  from Higgs mass considerations  in the large 
$\tan\beta$ regime.

%%%%%%%%%%%%%%%%%%%%%%%%%%%%%%%%%%%%%%%%%%%%%%%%%%%%%%%%%%%%%%%
%%%%%%%%%%%%%%%%%%%%%%%%%%%%%%%%%%%%%%%%%%%%%%%%%%%%%%%%%%%%%%%

Since the stop and chargino masses are fixed, 
we can determine the stop--chargino contribution to 
${\rm Br} (b \ra s \gamma)$~\cite{BSgamma}.   
It is important to remember though that the physical 
branching ratio depends on additional parameters 
which are not relevant for an interpretation of the HERA data.
For a specific choice of these parameters and 
$m_{\tilde{\chi}^+} = 190$ GeV, 
Fig.~\ref{fig:mh_bsg}($b$), 
shows the branching ratio 
as a function of the stop mixing angle. We 
set the first and second generation 
squark masses to 2 TeV and consider all four branches 
defined by the signs of $\mu$ and $\widetilde A_t$. 
The QCD scale has been fixed at $Q =  m_b/2$, a value
that reproduces the next-to leading-order 
corrections\footnote{A complete next-to leading-order computation
          in the supersymmetric model has not yet been performed.}
in the SM.
Setting $Q = m_b$ would typically reduce the rates 
by about $20\%$, giving a measure
of the theoretical uncertainties in these calculations.
For  $\tan\beta \simgt 10$, an acceptable
rate may be obtained only by large cancellations between  different
contributions to the amplitude, which demands
specific values of the Higgs and supersymmetric
mass parameters. 
For the above parameter choice and $\mu = - 1 \tev$, 
we find no solutions for $\tan\beta =10$, 50. 
Similarly, for $\mu = + 1 \tev$ and $\tan\beta = 50$,  
acceptable solutions may be found for only one
of the branches of $A_t$. 
Although this seems to prefer lower values of $\tan \beta$,
the situation can change for other parameter choices.
In fact, with suitable parameters, we can always reproduce the 
experimental results for ${\rm Br} (b \ra s \gamma)$
in the scenario under consideration.

In all of the above, we have held $m_D = 2 \tev$.
Although our scenario is consistent with lower
values of this mass parameter, the  analysis becomes more
involved in the large $\tan\beta$ region, where the mixing
in the sbottom sector becomes large. As noted earlier, 
a large value of $\theta_\sb$ will modify the relevant 
branching fractions, allowing for 
lower values of $\mu$ for a given $M_1 / M_2$ and $\tan \beta$.
On the other hand, since large $\theta_\sb$ generally implies 
a large $\theta_\st$, the atomic parity violation constraints 
become more stringent in this case. Observe that, for an appropriate
choice of the mass parameters,
larger values of the sbottom mixing angle may  allow for
a reduction of the value of $\Delta\rho$, even for moderately
large values of the stop mixing angle.

To summarize, then, we have investigated the possibility that both
the neutral and the charged current anomalies seen at HERA at
high $Q^2$ are consequences of the resonant production of a stop,
which decays to an $e^+ d$ final state to give the NC events and to 
$b \ch$ followed by cascade decays to give the CC events. 
Something of this nature
is indicated by the negative results of the Tevatron search for
leptoquarks/squarks in the 200 GeV mass range. Among the various
scenarios discussed, the CC events are best explained through the
decay $\ch \ra c \sb_L$, which is CKM-suppressed, but can be sizeable
in the limit of a somewhat large $|\mu|$ parameter, when the
competing decay channel is considerably reduced. The kinematic
distributions in this scenario explain the observed CC excess rather
well and the parameter space which supports this solution is
perfectly consistent with electroweak precision measurements at LEP
and with other constraints such as those originating from radiative
$B$-decays, atomic parity violation and Higgs searches 
at LEP. Although it may seem
premature to demand precise agreement with data for the CC
excess reported by H1 and ZEUS, 
our scenario makes firm predictions which should be 
testable in forthcoming runs 
of the existing high-energy accelerator facilities.

The authors would like to acknowledge discussions with 
G.~Altarelli, G.~Giudice, S.~Lola and P.~Zerwas 
and are grateful to Y. Sirois for specific information regarding the H1
observations. We thank P.~Chankowski for confirming our 
numerical results for $\Delta \rho_\st$. 

%%%%%%%%%%%%%%%%%%%%%%%%%%%%%%%%%%%%%%%%%%%%%%%%%%%%%%%%%%%%


\newpage
\begin{thebibliography}{99}

\def\bib{\bibitem}
%%%%%%%%%%%%%%%%%%%%%%%%%%%%%%%%%%%%%%%%%%%%%%%%%%%%%%%%%%%%%%%

% journal macros
%
\def\tp{these proceedings}
\def\ib#1,#2,#3{       {\em ibid.\/ }{\bf #1} (19#2) #3}
\def\ap#1,#2,#3{       {\em Ann.~Phys.~(NY)\/ }{\bf #1} (19#2) #3}
\def\appb#1,#2,#3{     {\em Acta Phys.\ Polon.\/ }{\bf B#1} (19#2) #3}
\def\cpc#1,#2,#3{      {\em Comput. Phys. Commun.\/ }{\bf #1} (19#2) #3}
\def\ijmp#1,#2,#3{     {\em Int.~J.~Mod.~Phys.\/ } {\bf A#1} (19#2) #3}
\def\mpl#1,#2,#3 {     {\em Mod.~Phys.~Lett.\/ } {\bf A#1} (19#2) #3}
\def\np#1,#2,#3{       {\em Nucl.~Phys.\/ }{\bf B#1} (19#2) #3}
\def\npps#1,#2,#3{     {\em Nucl.~Phys.~B (Proc.~Suppl.)\/ }
                             {\bf B#1} (19#2) #3}
\def\plb#1,#2,#3{      {\em Phys.~Lett.\/ }{\bf B#1} (19#2) #3}
\def\pr#1,#2,#3{       {\em Phys.~Rev.\/ }{\bf #1} (19#2) #3}
\def\prd#1,#2,#3{      {\em Phys.~Rev.\/ }{\bf D#1} (19#2) #3}
\def\prep#1,#2,#3{     {\em Phys.~Rep.\/ }{\bf #1} (19#2) #3}
\def\prl#1,#2,#3{      {\em Phys.~Rev.~Lett.\/ }{\bf #1} (19#2) #3}
\def\prog#1,#2,#3{     {\em Prog.~Theor.~Phys.\/ }{\bf #1} (19#2) #3}
\def\rmp#1,#2,#3{      {\em Rev.~Mod.~Phys.\/ }{\bf #1} (19#2) #3}
\def\sp#1,#2,#3{       {\em Sov.~Phys.-Usp.\/ }{\bf #1} (19#2) #3}
\def\zpc#1,#2,#3{      {\em Z. Phys.\/ }{\bf C#1} (19#2) #3}
%%%%%%%%%%%%%%%%%%%%%
\bib{H1} 
C.~Adloff \etal\ (H1 Collab.), \zpc74,97,{191}.

\bib{ZEUS} 
J.~Breitweg \etal\ (ZEUS Collab.),  \zpc74,97,{207}.

\bib{Adler}  
S.L.~Adler, Princeton preprint IASSNS-HEP-97-12,  hep-ph/9702378.

\bib{ChRa} 
D.~Choudhury and S.~Raychaudhuri, \plb401,97,{54}.

\bib{CERN_gang}G.~Altarelli, J.~Ellis, G.F.~Giudice, S.~Lola and M.L.~Mangano,
       CERN preprint CERN-TH-97-040, hep-ph/9703276.

\bib{DrMo}
H.~Dreiner and P.~Morawitz, hep-ph/9703279.

\bib{DESY_gang}J.~Kalinowski, R.~R\"uckl, H.~Spiesberger and P.M.~Zerwas, 
               \zpc74,97,{595}.

\bib{Blum} 
J.~Bl\"umlein, \zpc74,97,{605}.

\bib{BKMW} 
K.S.~Babu, C.~Kolda, J.~March-Russell and F.~Wilczek, 
          Princeton preprint IASSNS-HEP-97-04, hep-ph/9703299.

\bib{HeRi}
J.L.~Hewett and T.G.~Rizzo, SLAC preprint SLAC-PUB-7430,
hep-ph/9703337.

\bib{BCHZ}V.~Barger, K.~Cheung, K.~Hagiwara and 
D.~Zeppenfeld, Madison preprint 
         MADPH-97-991, hep-ph/9703311.

\bib{AlGiMa}
G.~Altarelli, G.F.~Giudice and M.L.~Mangano, CERN preprint
CERN-TH/97-101, hep-ph/9705287.

\bibitem{Low_ener} 
V.~Barger, G.F.~Giudice, and T.~Han, \prd40,89,{2987};\\
         G.~Bhattacharyya and D.~Choudhury, \mpl10,95,{1699}; \\
         G.~Bhattacharyya, J.~Ellis and K.~Sridhar, \mpl10,95,{1583};\\
         K.~Agashe and M.~Graesser, \prd54,96,{4445}.

\bib{Old_HERA} 
J.~Hewett, in {\em Snowmass Summer Study} (1990), p. 566;\\
T.~Kon and T.~Kobayashi, \plb270,91,{81};\\ 
T.~Kon, T.~Kobayashi and K.~Nakamura, in {\em Physics at HERA},
            Hamburg (1991), vol. 2, p.  1088;\\ 
J.~Butterworth and R.~Dreiner, \np397,93,{3};\\ 
T.~Kon, T.~Kobayashi, S.~Kitamura, K.~Nakamura 
              and S.~Adachi, \zpc61,94,{239};\\
T.~Kon, T.~Kobayashi and S.~Kitamura, \plb333,94,{263}; \\ 
T.~Kobayashi, S.~Kitamura and T. Kon, \ijmp11,96,{1875}; \\ 
E.~Perez, Y.~Sirois and H.~Dreiner, in {\em Future Physics at HERA},
            Hamburg (1996).

\bib{ElLoSr}
J.~Ellis, S.~Lola and K.~Sridhar, CERN preprint CERN-TH/97-109,
hep-ph/9705416.

\bib{JoRaVe}
A.S.~Joshipura, V.~Ravindran and S.K.~Vempati, Ahmedabad
preprint PRL-TH-97-016, hep-ph/9706482.

\bib{CDF_LQ}
H.S.~Kambara for the CDF Collaboration, CDF-PUB-EXOTIC-GROUP-4222, 
hep-ex/9706026, 
presented at the 12th
Workshop on Hadron Collider Physics (HCP 97), Stony Brook,

\bib{D0_LQ}
D.M.~Norman for the D0 Collaboration, 
FERMILAB-CONF-97-224-E, hep-ex/9706027, 
presented at the 12th Workshop
on Hadron Collider Physics (HCP 97), Stony Brook,

\bib{BaKoRu}
K.S.~Babu, C.~Kolda and J.~March-Russell, Princeton preprint
IASSNS-HEP-97-64, hep-ph/9705414.

\bib{GuRo}M.~Guchait and D.P.~Roy, Tata Inst. preprint TIFR-TH-97-29, 
         hep-ph/9707275.

\bib{ZEUS_new}D.~Krakauer (ZEUS Collaboration) presented at the 12th
     Workshop on Hadron Collider Physics (HCP 97), Stony Brook,

\bib{Kim}
J.E.~Kim and P.~Ko, Seoul preprint SNUTP~97-026, hep-ph/9706387.

\bib{KonNew}
T.~Kon, T.~Matsushita and T.~Kobayashi, Seikei preprint ITP-SU-97-03,
hep-ph/9707355.

\bib{Alta}
G.~Altarelli, \appb28,97,{991}.

\bib{sneut_LEP}
V.~Barger, W.Y.~Keung and R.J.N.~Phillips, \plb364,95,{27}.

\bib{CGLW}
M.~Carena, G.F.~Giudice, S.~Lola and C.E.M.~Wagner, \plb395,97,{225}.

\bib{4-jet}
D.~Buskulic \etal, (ALEPH Collab.) \zpc71,96,{179};\\
M.~Acciari \etal (L3 Collab.) CERN preprint CERN-PPE-97-057;\\
W.D.~Schlatter for the LEP Working Group on Four-Jet final states, 
CERN seminar, Feb. 25, 1997;\\
G.~Cowan for the ALEPH Collab., CERN seminar, Feb. 25, 1997. 

\bib{H1_old} 
See, \eg, S.~Aid \etal\ (H1 Collab.), \zpc71,96,{211}.

\bib{Cohen} A.G.~Cohen, D.B.~Kaplan and A.E.~Nelson, \plb388,96,{588}.

\bib{Sirois}
Y.~Sirois, talk presented on behalf of the H1 Collab. at CERN (March
4, 1997) and private communication.

\bib{Strumia} 
L.~Giusti and A.~Strumia, Pisa preprint IFUP-TH-23-97, hep-ph/9706298;\\
A.~Deandrea, Marseilles preprint CPT-97-P-3496, hep-ph/9705435.

\bib{QCD}
T.~Plehn, H.~Spiesberger, M.~Spira and P.M. Zerwas, \zpc74,97,{611}.

\bib{MRSH} 
A.D. Martin, R.G. Roberts and W.J. Stirling, \plb306,93,{145}; errat.
       \ib309,93,{492}.

\bib{PDFLIB} 
H.~Plothow-Besch, \cpc75,93,{396}.

\bib{Jac_Blo} 
S.~Bentvelsen, J.~Engelen and P.~Kooijman, Proc. Workshop on Physics
at HERA, ed. W.~Buchm\"uller and G.~Ingelman (DESY-Hamburg, 1991), p.
25;\\ 
K.C.~Hoeger, {\em ibid.}, p. 43.

\bib{KKK}T.~Kon, T.~Kobayashi and S.~Kitamura, \plb376,96,{227}.

\bib{Muon}T.~Ahmed \etal (H1 Collaboration), DESY preprint DESY-94-248 
      (1994).

\bib{APV} C.S.~Wood \etal, {\em Science} {\bf 275} (1997) 1759.

\bib{DrHa} 
M.~Drees and K.~Hagiwara \prd42,90,{1709}.

\bib{Higgs_mass} For the Higgs mass we use the two-loop 
    expressions given in\\
    M.~Carena, M.~Quir\'os and C.E.M.~Wagner, 
                     \np461,96,{407};\\
     H.E.~Haber, R.~Hempfling and A.H.~Hoang, CERN preprint CERN-TH/96-216, 
     hep-ph/9609331.

\bib{BSgamma} S.~Bertolini, F.~Borzumati, A.~Masiero and G.~Ridolfi,
                      \np353,91,{591};\\
              R.~Barbieri and G.~Giudice, \plb309,93,{86};\\
              M.S.~Alam \etal (CLEO Collaboration), \prl74,95,{2885}.

%%%%%%%%%%%%%%%%%%%%%%

\end{thebibliography}
\end{document}